**Symplectic connection third-order Hall effect in a room-temperature ferromagnet**

Yu Cao[1†], Xukun Feng[2†], Yiming Guo[1†], Huiying Liu[3], Qia Shen[1], Hongliang Chen[1], Wanxi Gong[1], Yu Yang[1], Dandan Guan[1], Yaoyi Li[1], Shiyong Wang[1], Hao Zheng[1], Canhua Liu[1], Xiaoxue Liu[1], Yumeng Yang[4], Xuepeng Qiu[5], Ruidan Zhong[1], Jinfeng Jia[1], Shengyuan A. Yang[6*], Cong Xiao[2*], Liang Liu[1*]

[1]*Tsung-Dao Lee Institute, Key Laboratory of Artificial Structures and Quantum Control (Ministry of Education), School of Physics and Astronomy, Shanghai Jiao Tong University, Shanghai 200240, China*

[2]*Interdisciplinary Center for Theoretical Physics and Information Sciences (ICTPIS), Fudan University, Shanghai 200433, China*

[3]*School of Physics, Beihang University, Beijing 100191, China*

[4]*Shanghai Engineering Research Center of Energy Efficient and Custom AI IC, School of Information Science and Technology, ShanghaiTech University, Shanghai 201210, China*

[5]*Shanghai Key Laboratory of Special Artificial Microstructure Materials and Technology and School of Physics Science and Engineering, Tongji University, Shanghai, 200092, China*

[6]*Research Laboratory for Quantum Materials, Department of Applied Physics, The Hong Kong Polytechnic University, Kowloon, Hong Kong, China*

[†]These authors contributed equally to this work.
*e-mail: shengyuan.yang@polyu.edu.hk; congxiao@fudan.edu.cn; liul21@sjtu.edu.cn

**Third-order nonlinear Hall effects (THE) have recently attracted considerable experimental interest as powerful probes for quantum geometric properties in emergent quantum materials, encompassing quadrupole moments of quantum metric[1-3] and Berry curvature[4,5]. Here, we report a fundamentally new THE in room-temperature van der Waals ferromagnet Fe₃GaTe₂ from second-order Berry connection polarizability[6], which manifests a higher-order characterization of band geometry called symplectic connection[7].**

**Our observations show that the third-order transverse response in Fe$_3$GaTe$_2$ is odd to magnetization, vanishes above the Curie temperature and remains independent of driving current directions. Scaling law analysis combined with first-principles calculations establishes this response as the symplectic-connection-induced THE. This discovery opens the door to probing high-order quantum geometric properties beyond Berry curvature and quantum metric through nonlinear transport, unveiling the potential of exploring nonlinear Hall phenomena in broad classes of magnets without breaking inversion symmetry. Moreover, the room-temperature manipulation of THE holds promises for device applications based on harnessing the quantum-geometric connection structure.**

Nonlinear Hall effects provide a powerful tool for probing quantum geometry in solids. The second-order Hall effects reveal the momentum-space dipole of Berry curvature[8-11] in non-magnetic materials with inversion symmetry ($\mathcal{P}$) breaking as well as the dipole of quantum metric[12-15] in $\mathcal{P}$-broken magnets. Following this paradigm, the third-order nonlinear Hall effect (THE) should be induced by the quadrupole moments of quantum metric and of Berry curvature in non-magnetic and magnetic materials, respectively[16,17]. Experiments have indeed probed the quantum metric quadrupole in MoTe$_2$[1] and WTe$_2$[2] among other non-magnetic materials[3,18-20], and very recently also the Berry curvature quadrupole in antiferromagnetic FeSn[4] and MnBi$_2$Te$_4$[5]. On the other hand, a fundamentally new THE was recently proposed by theory[6], which does *not* share quantum geometric origins with second-order transport but rather is determined by a novel geometric object called second-order Berry connection polarizability (SBCP). The SBCP quantifies the second-order polarizability of Berry connection to electric field and hence geometrically represents the positional shift of an electron wave packet in response to the second-order electric field: $\delta r_a = T_{abc} E_b E_c$ (summation over repeated indices $a, b, c$ for spatial directions). The effective position in the presence of SBCP is thus shifted as $r_a \to i\partial_a + \delta r_a$ ($\partial_a \equiv \partial/\partial k_a$ with $k_a$ the wave vector) and becomes non-canonical[21,22]: $[r_a, r_b] = i(\partial_a T_{bcd} - \partial_b T_{acd}) E_c E_d$. This non-canonicity leads to an anomalous velocity[21] felt by the electron wave packet $\delta v_a = -i[r_a, eE_b r_b] = e(\partial_a T_{bcd} - \partial_b T_{acd}) E_b E_c E_d$, which, upon summing over occupied states, underlies a third-order transverse current $\sim \int_{occ.} \partial T$ that is quantified by the dipole of SBCP (as shown in Fig. 1). This effect has been proposed for an antiferromagnetic two-band model[6], but is yet to be realized in experiments and explored in real materials. Moreover, in the Riemannian geometry language[7,23] introduced by Ahn *et al.* for electronic Bloch states, the SBCP has a Riemannian geometric origin in the symplectic connection[24]. This is a higher-order characterization of Riemannian structure compared to the quantum geometric tensor that is related to Berry curvature and quantum metric. While the latter two geometric objects have been extensively studied in anomalous and nonlinear Hall effects, the way the symplectic connection regulates electronic transport remains undetected.

In contrast to the dipoles of Berry curvature and of quantum metric, both of which require breaking $\mathcal{P}$, the dipole of SBCP is compatible with $\mathcal{P}$ (Fig. 1). As an intrinsic response without relying on electronic relaxation processes, the time-reversal symmetry $\mathcal{T}$ must be broken in the band structures to support this effect. Then, given that ferromagnetic metals are usually centrosymmetric and the scarcity of room-temperature antiferromagnetic metals, we focus on ferromagnetic metals and identify $Fe_3GaTe_2$ as a suitable platform. As a newly observed room-temperature van der Waals ferromagnet[25], $Fe_3GaTe_2$ has been the subject of extensive research due to its potential applications in magnetic tunnel junctions[26], spin-orbit torque devices[27-29], skyrmions-based devices[30-32], and unconventional transport behaviors[33,34]. Its magnetic symmetry (see below) supports THE, but to date nonlinear electrical transport of $Fe_3GaTe_2$ is still unexplored. Here, we report the experimental observation of THE up to room temperature in $Fe_3GaTe_2$, which is odd to magnetization (*M*) and only observable in the ferromagnetic state, disappearing in the paramagnetic state. The effect exhibits an isotropy with respect to the current flowing direction in the $Fe_3GaTe_2$ (0001) plane, enabling the isolation of individual element of nonlinear conductivity tensor, which is prerequisite for quantitative experimental exploration of underlying physical mechanisms. Our scaling law analysis and first-principles calculations point to the existence of SBCP-dipole induced THE and provide a room-temperature probe of the symplectic connection.

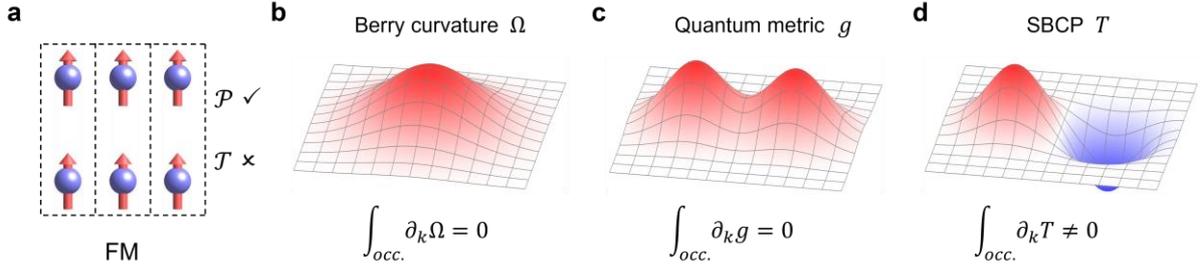

**Figure 1| Schematics of space-time symmetry constraints on nonlinear transport induced by momentum-space dipole of different quantum geometric quantities. a,** Schematics of ferromagnets with inversion symmetry. **b,** The Berry curvature distribution in $k$-space is even, leading to vanishing Berry curvature dipole integrated over occupied states (occ.). **c,** The even distribution of quantum metric and vanishing quantum metric dipole. **d,** The SBCP has an odd distribution in $k$-space, resulting in a finite SBCP dipole.

**Basic characterization of Fe$_3$GaTe$_2$**

Figures 2a and 2b show the side and top views of the Fe$_3$GaTe$_2$ crystal structure. Bulk Fe$_3$GaTe$_2$ crystallizes in the P6$_3$/$mm'c'$ magnetic space group, featuring a sixfold screw axis symmetry and multiple generalized mirror symmetries. These include glide magnetic mirror symmetry ($m \cdot c_{1/2} \cdot \mathcal{T}$) and magnetic mirror symmetry ($m \cdot \mathcal{T}$), both aligned parallel to the *c*-axis (Fig. 2b). Additionally, mirror symmetry ($m$) is preserved within Ga atomic planes in Fe$_3$GaTe$_2$ monolayer, as illustrated in Fig. 2a. Above the Curie temperature ($T_C$), bulk Fe$_3$GaTe$_2$ retains both $\mathcal{P}$ and $\mathcal{T}$ symmetries within the 6/$mmm$ point group. Below $T_C$, Fe$_3$GaTe$_2$ exhibits perpendicular magnetization anisotropy (PMA) with a magnetic point group of 6/$mm'm'$. Supplementary Figure S1 presents a high-angle annular dark field-scanning transmission electron microscopy (HADDF-STEM) image of the studied Fe$_3$GaTe$_2$ crystal, confirming its van der Waals layered structure. For electrical transport measurement, we exfoliated a Fe$_3$GaTe$_2$ flake onto a SiO$_2$/Si substrate and fabricated it into a circular disc with a thickness (*t*) of ~77 nm and a diameter (*L*) of 50 μm, as shown in Fig. 2c (see Method for the details). Ti/Au electrodes were deposited via e-beam evaporation after a standard photolithography process. Supplementary Fig. S2a shows the temperature (*T*) dependence of longitudinal resistance ($R_∥$) of our Fe$_3$GaTe$_2$ sample, exhibiting characteristic metallic behavior. Supplementary Figure S2b and its inset show the derived resistivity ($\rho_\parallel$) and its first derivative ($d\rho_\parallel/dT$) as a function of $T$, where the peak of $d\rho_\parallel/dT$ indicates a $T_C$ of 350 K. Supplementary Fig. S3 shows the anomalous Hall resistance ($R_\perp$) as a function of the out-of-plane magnetic field ($H_z$), i.e., the anomalous Hall effect (AHE) loops, at temperatures ($T$) ranging from 10 K to 400 K. Nearly square AHE loops were observed at 300 K and below confirming good PMA, while the 310-350 K shows zero remanence despite the persistent ferromagnetic order.

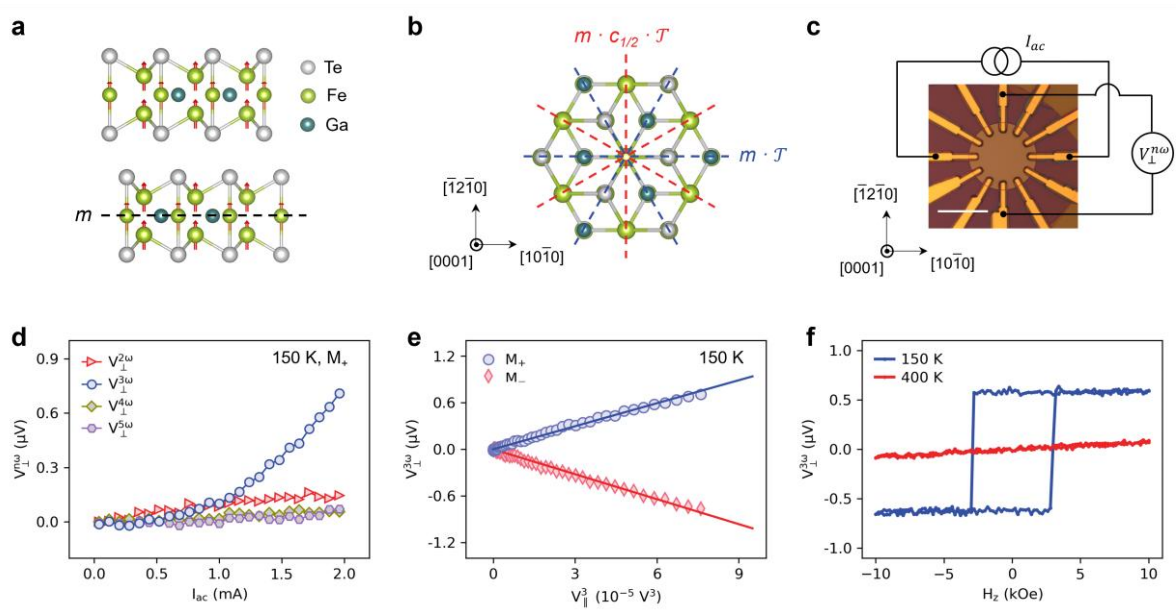

**Figure 2| Crystal symmetry and nonlinear transport of Fe$_3$GaTe$_2$. a,** Side view of the magnetic structure of bilayer Fe$_3$GaTe$_2$. **b,** Plane view of Fe$_3$GaTe$_2$. $m$, $c_{1/2}$, and $\mathcal{T}$ represent the mirror operation, half-translation operation (translation direction along [0001]) and time-reversal operation, respectively. $m \cdot \mathcal{T}$ and $m \cdot c_{1/2} \cdot \mathcal{T}$ represent the magnetic mirror symmetry (blue) and glide magnetic mirror symmetry (red), respectively. **c,** Schematic of the Hall measurement configuration of the Fe$_3$GaTe$_2$ device. The circular disc is Fe$_3$GaTe$_2$ flake after etching and the yellow bars are Ti/Au electrodes. The scale bar is 50 μm. **d,** Current dependences of 2$^{nd}$ to 5$^{th}$ order transverse responses measured at 150 K for $M+$ state. **e,** Scaling between the third-order transverse voltage $V_\perp^{3\omega}$ and the cubic of the longitudinal voltage $V_\parallel^3$ for $M+$ (blue) and $M-$ (red) states measured at 150 K. **f,** $V_\perp^{3\omega}$ as a function of the out-of-plane magnetic field ($H_z$) at 150 K (blue) and 400 K (red). A fixed ac current of 1.8 mA is applied for this measurement.

**Nonlinear transport measurements of $\mathcal{T}$-odd and isotropic THE**

We then characterized the nonlinear transport of Fe$_3$GaTe$_2$. As shown in Fig. 2c, an ac current ($I_{ac}$) was applied along the $[10\bar{1}0]$ direction of Fe$_3$GaTe$_2$, and the transverse voltage at different orders ($V_\perp^{n\omega}$) were measured using a standard lock-in technique (see Methods). First, we measured $V_\perp^{n\omega}$ with $n$ ranging from 2 to 5 at 150 K (below $T_C$), as shown in Fig. 2d. Before the measurement, an out-of-plane magnetic field ($H_z$) higher than the coercive field was applied to initialize the magnetization to the +z direction. Then $H_z$ was swept to 0, and the magnetization ($M$) remained nearly unchanged due to the PMA of the material. This state is denoted as "$M+$", while "$M-$" represents the state initialized in the -z direction. The third-order signal ($V_\perp^{3\omega}$) was found to be the most prominent among all nonlinear signals (Fig. 2d). The small second-order signal ($V_\perp^{2\omega}$) likely originates from the second-order nonlinear Hall effect caused by slight inversion-symmetry breaking by surfaces[1]. Figure 2e shows the relationship between $V_\perp^{3\omega}$ and the cubic of the longitudinal voltage ($V_\parallel^3$) for $M+$ (blue) and $M-$ (red) states at 150 K, where $V_\parallel$ is calculated by $I_{ac}R_\parallel$. The $V_\perp^{3\omega} - V_\parallel^3$ relation exhibits good linearity for both magnetization states. $V_\perp^{3\omega}$ reverses sign upon magnetization reversal from $M+$ to $M-$, confirming its odd symmetry to $M$.

We further investigated $V_\perp^{3\omega}$ as a function of $H_z$ at two representative temperatures (150 K and 400 K), above and below $T_C$, under a fixed ac current of 1.8 mA. At 150 K (ferromagnetic state), the $V_\perp^{3\omega} - H_z$ relation displays a square hysteresis loop with the same coercive field as the AHE loop shown in Supplementary Fig. S3a. At 400 K (paramagnetic state), $V_\perp^{3\omega}$ vanishes at zero magnetic field, but can only be driven linearly by the field as a magneto-nonlinear transport effect[35,36]. Frequency-dependent measurements confirmed that $V_\perp^{3\omega}$ remains unchanged across different frequencies (Supplementary Fig. S4) and Joule heating effect on the observed signal was also excluded (Supplementary Section 5). These results together demonstrate that the observed triple-frequency transverse response at zero magnetic field is a $\mathcal{T}$-odd THE correlated with the magnetic order of the sample. The amazing absence of $\mathcal{T}$-even THE in our observation is in effect a consequence of symmetry. Bulk Fe$_3$GaTe$_2$, with its 6/$mmm$ point group, strictly prohibits the $\mathcal{T}$-even THE in the (0001) plane[16]. This consistency between theory and experiment further provides compelling evidence that our observation manifests the intrinsic property of bulk Fe₃GaTe₂.

The magnetic group symmetry also imposes stringent constraints on the $\mathcal{T}$-odd THE. Remarkably, it enforces that the third-order transverse nonlinear signal $E_\perp^{3\omega}/E_\parallel^3$ is dictated by only one independent element of the third-order nonlinear conductivity (see Methods for derivations):

$$\frac{\sigma E_\perp^{3\omega}}{E_\parallel^3} = \chi_{yxxx}, \tag{1}$$

which thus remains intact upon varying the driving current direction. Here, $\sigma$ is the Drude conductivity, which is isotropic as required by the 6/*mmm* symmetry. This isotropy of THE is not shared by previously reported THE[1-3,18,20]. It has the advantage of isolating individual nonlinear conductivity element, which is an experimental prerequisite for quantitative investigation of underlying physical mechanisms. It also allows for feasible but accurate determination of nonlinear conductivity without invoking careful alignment between electrodes and crystal axes.

To verify Eq (1), we performed angular-dependent measurements of THE. We applied an electrical current at various angles ($\theta_I$) relative to the [10$\bar{1}$0] axis within the (0001) plane of Fe₃GaTe₂ and subsequently measured the transport properties, as depicted in Fig. 3a. Figure 3b shows the $\theta_I$ dependence of longitudinal resistivity $\rho_\parallel$ measured at 150 K, where we observed an isotropic behavior. This confirms the absence of electrode misalignment or sample inhomogeneity in our Fe₃GaTe₂ device (see Supplementary Section 5 for details). Figure 3c shows $V_\perp^{3\omega}$ as a function of $I_{ac}$ at 150 K for different $\theta_I$ ranging from 0 to 150°. The $V_\perp^{3\omega}$-$I_{ac}$ relation shows negligible angular dependence, with all data points following the linear scaling between $V_\perp^{3\omega}$ and $V_\parallel^3$ (Fig. 3c). We plot the derived slope $E_\perp^{3\omega}/E_\parallel^3 = L^2 V_\perp^{3\omega}/V_\parallel^3$ (where $L$=50 μm) for different $\theta_I$ ranging from 0 to 330° in Fig. 3d, comparing both *M+* (blue) and *M-* (red) states. These results further confirm the isotropic characteristic predicted by Eq. (1). When increasing the temperature to 400 K (paramagnetic state), $V_\perp^{3\omega}$ vanishes for all $\theta_I$ (Supplementary Fig. S6).

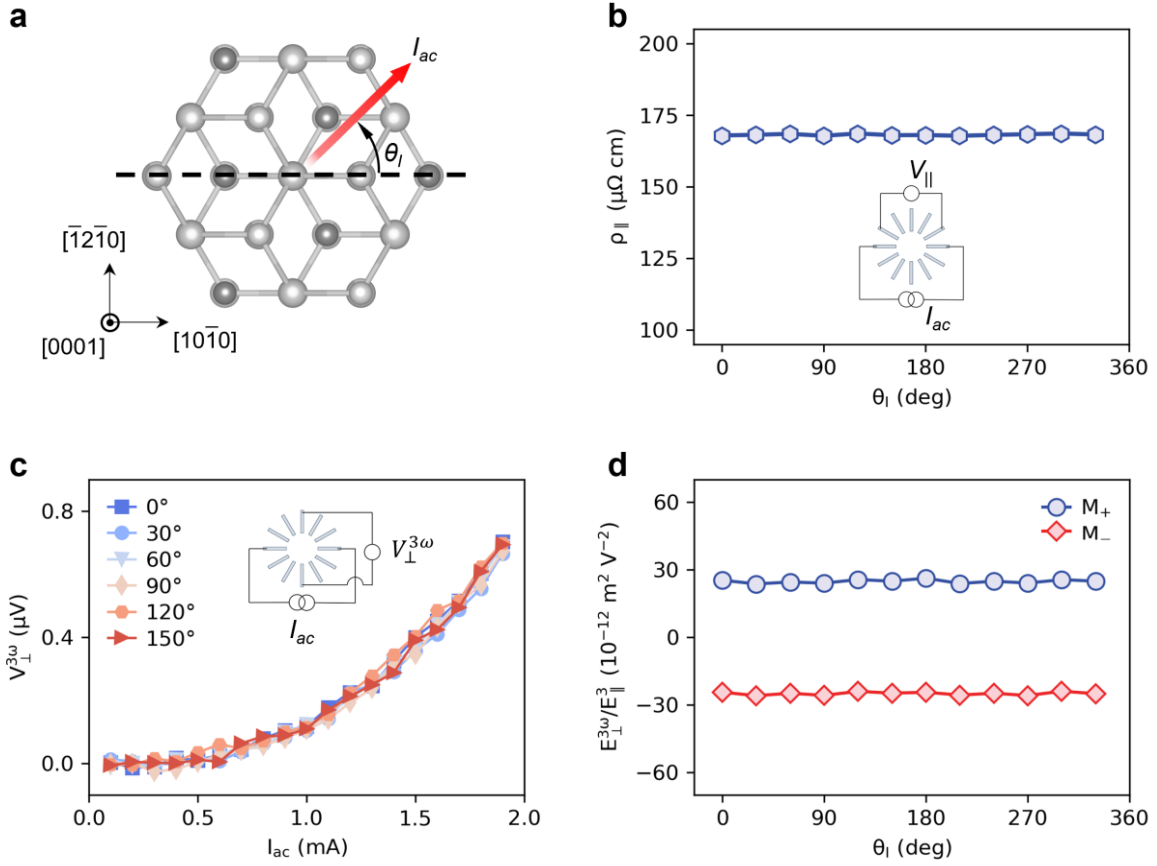

**Figure 3| Isotropy of third-order nonlinear transverse response in Fe₃GaTe₂. a,** Schematic of the angular-dependent measurement. The Fe₃GaTe₂ (0001) plane is grey-colored here for better visualizing the electric current. $\theta_I$ denotes the angle between the current direction and the $[10\bar{1}0]$ direction. **b,** $\theta_I$ dependence of longitudinal resistivity $\rho_\parallel$ at 150 K. Inset: measurement scheme for longitudinal resistivity measurement. **c,** $V_\perp^{3\omega}$ versus $I_{ac}$ for varying $\theta_I$ at 150 K. Inset: measurement scheme for transverse measurement. **d,** $\theta_I$ dependence of $\frac{E_\perp^{3\omega}}{E_\parallel^3}$ at 150 K for *M+* (blue) and *M-* (red) states.

**Physical origins: SBCP and symplectic connection**

To explore the origin of the observed THE, we examined the temperature dependence of $V_\perp^{3\omega}$. Supplementary Fig. S7a shows the $V_\perp^{3\omega}$-$I_{ac}$ relation at different *T* ranging from 110 K to 400 K. Accordingly, we derived the $V_\perp^{3\omega}$-$V_\parallel^3$ dependence as shown in Fig. 4a, which maintains good linearity between 100 K to 300 K before disappearing above 300 K. Figure 4b and Figure 4c present the temperature dependence of the conductivity ($\sigma=1/\rho_\parallel$) and the normalized third-order transverse nonlinear conductivity $\frac{\sigma E_\perp^{3\omega}}{E_\parallel^3}$. For the *M+* state (blue points), $\frac{\sigma E_\perp^{3\omega}}{E_\parallel^3}$ decreases with increasing temperature from 110 K to 300 K. The behavior for *M-* state (red points) is consistent with the $\mathcal{T}$-odd nature of the effect. As the temperature approaches 310 K, $\frac{\sigma E_\perp^{3\omega}}{E_\parallel^3}$ undergoes an abrupt transition to zero. This feature is analogous to the temperature dependence of the anomalous Hall resistance $R_\perp$ (Supplementary Fig. S7d). In the temperature range from 310 K to 350 K, both $\frac{\sigma E_\perp^{3\omega}}{E_\parallel^3}$ and $R_\perp$ vanish completely, indicating that out-of-plane magnetization is a prerequisite for the observation of $\mathcal{T}$-odd signals.

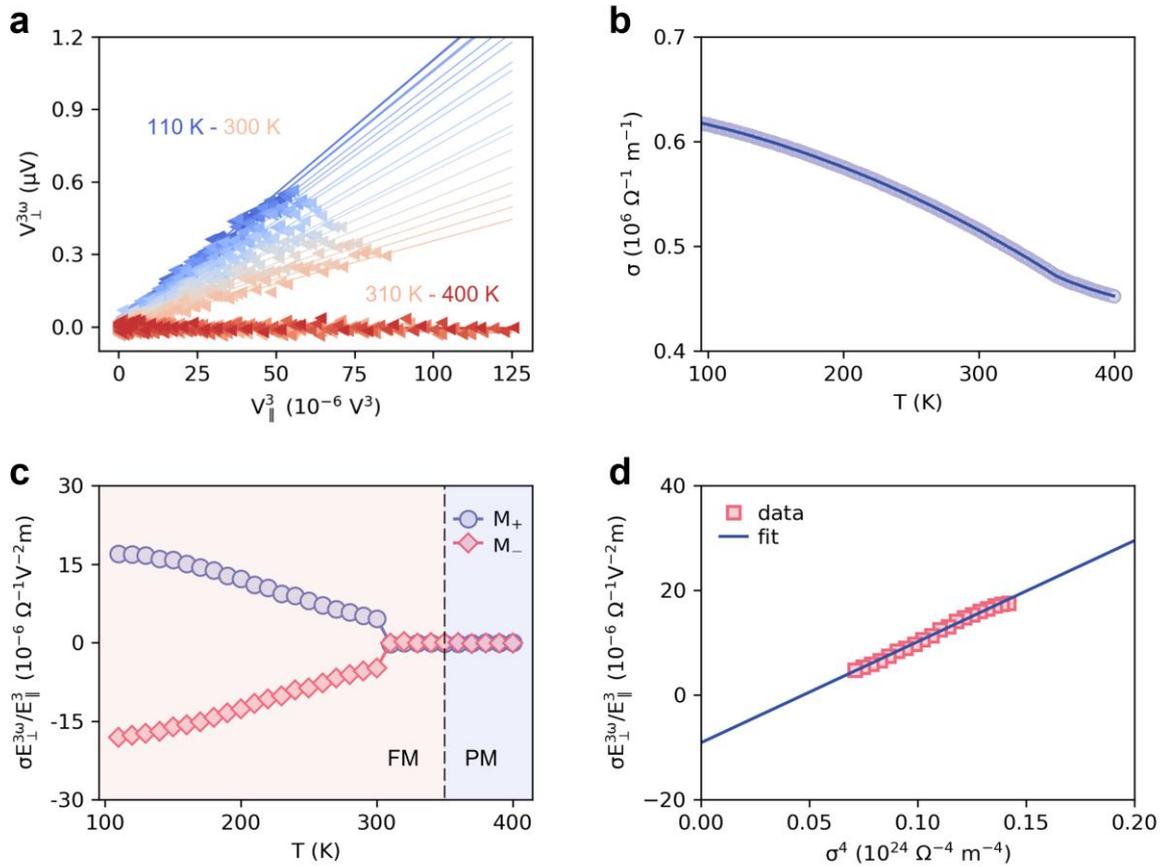

**Figure 4| Temperature dependence of third-order transverse response in Fe₃GaTe₂. a,** Scaling relation between the third-order transverse voltage $V_\perp^{3\omega}$ and the cubic of longitudinal voltage $V_\parallel^3$. **b,** Temperature dependence of electrical conductivity ($\sigma$). **c,** Normalized third-order transverse nonlinear conductivity $\frac{\sigma E_\perp^{3\omega}}{E_\parallel^3}$ versus temperature for *M+* (blue) and *M-* (red) states, with FM (ferromagnetic) and PM (paramagnetic) regimes indicated. **d,** Scaling relation between $\frac{\sigma E_\perp^{3\omega}}{E_\parallel^3}$ and the quartic of Drude conductivity $\sigma^4$.

Table 1 compares the THEs of Fe$_3$GaTe$_2$ and other materials that were measured at room temperature under zero magnetic field. Compared to TaIrTe$_4$[3] and VSe$_2$[18], Fe$_3$GaTe$_2$ exhibits a larger room-temperature transverse nonlinear signal and nonlinear conductivity. It is expected that the sizable room-temperature field-free third-order response will be useful for future (opto)electronic devices.

**Table 1. Comparison of zero-field room-temperature THE ever reported.**

| Materials | $E_\perp^{3\omega}/E_\parallel^3$ (10$^{-12}$m$^2$V$^{-2}$) | $\sigma E_\perp^{3\omega}/E_\parallel^3$ (10$^{-6}\Omega^{-1}$V$^{-2}$m) | $B$ (T) | $T$ (K) | Refs. |
|---|---|---|---|---|---|
| VSe$_2$ | <1.5 | 1.5 | 0 | 300 | 18 |
| TaIrTe$_4$ | ~0.8 | 0.7 | 0 | 300 | 3 |
| Fe$_3$GaTe$_2$ | 8 | 4.1 | 0 | 300 | This work |

The temperature-dependent data enable us to explore the scaling law of the third-order nonlinear conductivity in terms of the Drude conductivity, which has served as a pivotal tool for understanding experiments on THE[1-5,18-20]. As shown in Fig. 4d, a quartic scaling

$$\frac{\sigma E_\perp^{3\omega}}{E_\parallel^3} = \xi\sigma^4 + \eta \qquad (2)$$

with a prominent $\sigma$-independent intercept $\eta$ was observed in ferromagnetic Fe$_3$GaTe$_2$. This scaling law was also repeated in another device (Supplementary Fig. S8). The quartic term has only been reported recently in the canted state of antiferromagnet FeSn[4], and was attributed to the extrinsic mechanism of skew scattering[4]. In fact, this term may also be related to higher-order extrinsic processes, such as the anomalous second-order skew scattering, which is the combination of Berry curvature anomalous velocity and second-order skew scattering[37] (compositions of two skew scatterings). Up to now, these highly extrinsic terms are still hard to calculate in real materials. On the other hand, the $\sigma$-independent third-order transverse conductivity $\eta$ has not been reported before, and is the focus here. More importantly, the only known mechanism for this term is the intrinsic THE[6] induced by SBCP, which can be evaluated quantitatively for real materials.

The intrinsic third-order Hall conductivity of Fe$_3$GaTe$_2$ is estimated from[6] $\chi_{abcd}^{int} = -e^2 \int_k \Lambda_{abcd}$, which measures the anti-symmetrized SBCP dipole of the occupied states $\Lambda_{abcd} = \sum_n (\partial_b T^n_{acd} - \partial_a T^n_{bcd}) f_0$. Here, $f_0$ is the equilibrium Fermi-Dirac distribution, and $T^n_{abc}$ is the SBCP for band $n$ (see Methods for details). Figure 5a shows the calculated band structures with spin-orbit coupling, from which we obtain $\chi^{int}_{yxxx}$ of $-4.6\times10^{-6}$ $\Omega^{-1}$ V$^{-2}$ m at the intrinsic Fermi level, as shown in Fig. 5b. This calculation result agrees well with the experimental value of $-9.13 \times10^{-6}$ $\Omega^{-1}$V$^{-2}$m (estimated from Fig. 4d). Moreover, the SBCP can be decomposed into $T^n_{abc} = \sum_{m\neq n}(\Gamma^{nm}_{bac} + \Gamma^{nm}_{acb} + \Gamma^{nm}_{cba})/\omega^2_{nm} + TBC$. The first term is dictated by the symplectic connection[7,23,24] $\Gamma^{nm}_{bac} = -\Im(A^{mn}_b \mathfrak{D}^{nm}_a A^{nm}_c)$, with $A^{nm}_c$ being the interband Berry connection and $\mathfrak{D}^{nm}_a = \partial_a - i(A^n_a - A^n_m)$ the covariant derivative, while the second term is a three-band contribution $TBC = \Re \sum'_{m,\ell} \frac{A^{nm}_b A^{m\ell}_a A^{\ell n}_c + A^{nm}_a A^{m\ell}_c A^{\ell n}_b + A^{nm}_c A^{m\ell}_b A^{\ell n}_a}{\omega_{nm}\omega_{nl}}$, where $\omega_{nm}$ is the energy difference of two bands at the same $k$ and all the band indices cannot be equal in the summation. As shown in Fig. 5a, the band structures around the Fermi level of Fe$_3$GaTe$_2$ display multiple near-degeneracy regions, at which only two closest bands dominate the interband coherence response. In addition, in Fig. 5c, we plotted the distribution of SBCP dipole on the Fermi surface (via an integration by parts) $\sum_n (v_x T^n_{yxx} - v_y T^n_{xxx}) f'_0$ on the $k_z = 0$ plane, and found that large values are concentrated around such near-degeneracy regions. Therefore, the TBC, which must involve a relatively remote band, should be negligible compared to the symplectic connection term. We have performed concrete calculations of the TBC and found that it is indeed nearly three orders of magnitude smaller than the symplectic connection contribution. As such, the SBCP-dipole induced THE is dictated by the symplectic connection, as shown in Fig. 5b.

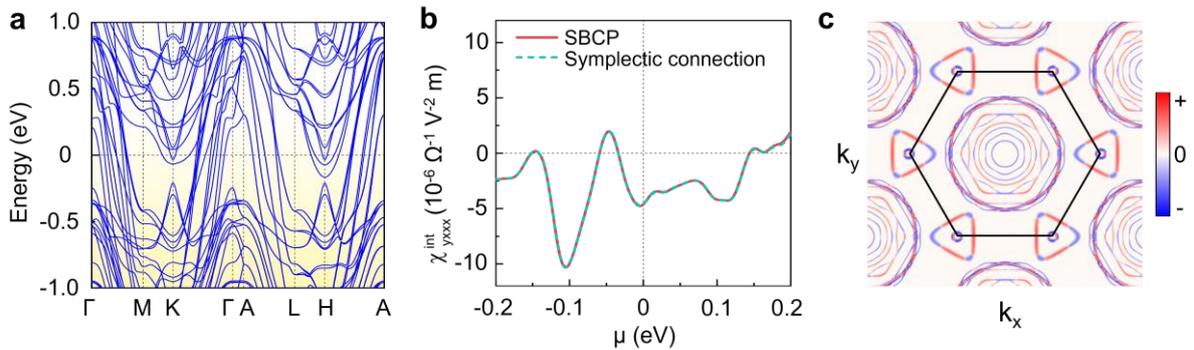

**Figure 5| Symplectic connection as the quantum geometric origin. a,** Calculated band structures of bulk ferromagnetic Fe$_3$GaTe$_2$ with spin-orbit coupling. **b,** Calculated third-order

intrinsic Hall conductivity $\chi_{yxxx}^{int}$ versus chemical potential. The dashed line represents the symplectic connection contribution, which totally dominates the SBCP response. **c,** $k$-resolved SBCP dipole on the Fermi surface $\sum_n (v_x T_{yxx}^n - v_y T_{xxx}^n) f_0'$ on $k_z = 0$ plane.

**Discussion and outlook**

The realization of the intrinsic THE in $Fe_3GaTe_2$ provides the first example in electronic transport of probing high-order quantum geometric properties beyond Berry curvature and quantum metric. Moreover, unlike all other THE mechanisms that are dependent on scattering time, the intrinsic THE is solely dictated by the band structures thus renders an accurate and robust probe of inherent material properties. In low-frequency nonlinear transport, the THE can be used also for rectification by applying an additional dc electric field along with a second-order ac field, as has been shown for nonmagnetic materials[19]. By this approach, in $Fe_3GaTe_2$ the revealed symplectic connection provides a new intrinsic mechanism for rectification.

The observed THE in $Fe_3GaTe_2$ also provides a rare example of nonlinear transport in ferromagnets, demonstrating the potential of nonlinear electronics in magnetic materials without breaking inversion symmetry. According to symmetry analysis, there are 19 centrosymmetric magnetic point groups that are compatible with the intrinsic THE (Methods), extending greatly the material classes for intrinsic rectification and rendering a broad platform for studying SBCP dipole and symplectic connection. In particular, all centrosymmetric ferromagnets support the intrinsic THE. This prospect may stimulate further theoretical and experimental investigations of nonlinear device functionalities enabled by higher-order quantum geometric properties in a wide variety of centrosymmetric ferromagnets which were deemed as useless in nonlinear electronics. Furthermore, new horizons will appear in extending intrinsic THE into nonmagnetic materials and few-layered materials, where the symplectic-connection structure can respectively be induced or manipulated by the external magnetic field[36] and gate field.

The revealed symplectic connection also play a role in other nonlinear transport and response phenomena[38,39]. For example, it should be linked to $\mathcal{T}$-even fourth-order nonlinear transport, which results from the accumulation of symplectic connection on the shifted Fermi surface by first-order electric field. Besides, in higher-order magnetoelectric response, the magnetization dependence of second-order electric polarization (SBCP) given by the symplectic connection yields a nonlinear magnetoelectric coefficient, which can lead to a nonlinear spin-orbit torque in magnetic insulators, providing a new route to low-consumption insulator spintronics. Moreover, the finding of symplectic connection in THE opens the door to discovering abundant high-order quantum geometric properties by nonlinear transport, such as metric connection, Riemann curvature, and symplectic curvature[7]. For example, the $\mathcal{T}$-odd intrinsic fourth-order nonlinear Hall effect should be related to a third-order Berry connection polarizability, whose connection to Riemann curvature and symplectic curvature is an interesting future question. All these topics render numerous opportunities for looking into unexplored connections between high-order quantum geometric structures of electronic states and macroscopic material properties.

## Methods

### Sample preparation and device fabrication

Single-crystalline $Fe_3GaTe_2$ nanoflakes were mechanically exfoliated from bulk crystals using Scotch tape and transferred onto $SiO_2$/Si substrate with the assistance of polydimethylsiloxane (PDMS). Selected flakes were patterned into circular disc geometries (50 μm diameter) using laser direct-write lithography and argon ion milling. Twelve Ti/Au (10 nm/80 nm) electrodes were deposited by electron-beam evaporation following lithographic patterning.

### Scanning transmission electron microscope

Cross-sectional TEM specimens of $Fe_3GaTe_2$ were prepared using a ZEISS Crossbeam 350 focused ion beam system and analyzed using a ThermoFisher Spectra 300 STEM equipped with an X-CFEG electron source, an S-CORR probe aberration corrector, operating at an accelerating voltage of 300 kV. The High-angle annular dark-field-STEM images were acquired with a probe-forming angle of 30 mrad and a collection angle of 46-200 mrad.

### Electrical transport measurement

The electrical transport properties of the devices were characterized using a Quantum Design Physical Property Measurement System (PPMS). An ac current with a frequency of 317.3 Hz was supplied by a Keithley 6221 current source, while the 1st to 5th-order transverse voltage signals were measured using lock-in amplifiers (Stanford Research Systems SR830).

### Angular dependence of THE

The third-order charge current can be expressed in terms of the third-order nonlinear conductivity $\chi$ as $j_a = \chi_{abcd} E_b E_c E_d$, where the Einstein summation convention is implied over Cartesian indices $a, b, c, d \in \{x,y\}$. The magnetic point group of ferromagnetic bulk $Fe_3GaTe_2$ is $6/mm'm'$, whose generators are $\mathcal{P}$, $C_{2z}$, $C_{3z}$ and $C_{2x}\mathcal{T}$. Dictated by this group, the permittable elements of $\mathcal{T}$-odd $\chi_{abcd}$ are $\chi_{yxxx} = -\chi_{xyyy}$, $\chi_{yxyy} = -\chi_{xyxx}$, $\chi_{yyxy} = -\chi_{xxyx}$, and $\chi_{yxxx} = \chi_{yxyy} + \chi_{yyxy} + \chi_{yyyx} = -(\chi_{xxxy} + \chi_{xxyx} + \chi_{xyxx})$. With the applied in-plane electric field $\boldsymbol{E}_\parallel = E(\cos\theta, \sin\theta)$, where $E_x = E\cos\theta$ and $E_y = E\sin\theta$, the third-order Hall current density can be expressed as

$$j^{(3)} = \begin{bmatrix}(\chi_{xxxy} + \chi_{xxyx} + \chi_{xyxx})E_x^2 E_y + \chi_{xyyy}E_y^3 \\ (\chi_{yxyy} + \chi_{yyxy} + \chi_{yyyx})E_x E_y^2 + \chi_{yxxx}E_x^3\end{bmatrix}^T \begin{bmatrix}-\sin\theta \\ \cos\theta\end{bmatrix} = \chi_{yxxx}E_\parallel^3.$$

Meanwhile, it can also be expressed as $j^{(3)} = \sigma E_\perp^{(3)}$, with $\sigma$ being the isotropic longitudinal Drude conductivity for Fe$_3$GaTe$_2$. These two equations yield $\sigma E_\perp^{(3)} = \chi_{yxxx}E_\parallel^3$, i.e., Eq. (1).

**Intrinsic THE formulae in terms of SBCP and symplectic connection**

The intrinsic third-order nonlinear Hall conductivity is given by

$$\chi_{abcd}^{int} = -e^2 \int [d\boldsymbol{k}] \Lambda_{abcd},$$

where $[d\boldsymbol{k}]$ is the the short-hand notation of $d^3\boldsymbol{k}/(2\pi)^3$, and

$$\Lambda_{abcd} = \sum_n (\partial_b T_{acd}^n - \partial_a T_{bcd}^n) f_0$$

is the anti-symmetrized SBCP dipole of the occupied states. Via an integration by parts, we have

$$\chi_{abcd}^{int} = -e^2 \int [d\boldsymbol{k}] \sum_n (v_a T_{bcd}^n - v_b T_{acd}^n) f_0',$$

where $f_0'$ is the energy derivative of the Fermi-Dirac distribution. The SBCP of band $n$ is expressed as (set $\hbar = 1$)

$$T_{abc}^n = \Re \sum_m{}' (\mathcal{U}_{abc}^{nm} + \mathcal{U}_{bac}^{nm} - \mathcal{U}_{bca}^{mn}) + TBC,$$

where

$$\mathcal{U}_{abc}^{nm} = \frac{iA_b^{nm}\mathfrak{D}_a^{nm}A_c^{mn}}{\omega_{nm}^2} - \frac{i\Delta_a^{nm}}{\omega_{nm}^3}A_b^{nm}A_c^{mn} = \frac{i}{\omega_{nm}^2}C_{bac}^{nm} - \frac{i\Delta_a^{nm}}{\omega_{nm}^3}\mathcal{Q}_{bc}^{nm},$$

and the three-band contribution (TBC) reads

$$TBC = \Re \sum_{m,\ell}{}' \frac{2A_a^{nm}A_b^{m\ell}A_c^{\ell n} + A_b^{nm}A_a^{m\ell}A_c^{\ell n}}{\omega_{nm}\omega_{nl}}.$$

Here $\mathfrak{D}_a^{nm} = \partial_a - i(A_a^n - A_a^m)$ is the covariant derivative, $\Delta_a^{nm} = \partial_a\omega_{nm}$, $\mathcal{Q}_{bc}^{nm} = A_b^{nm}A_c^{mn}$ is the Hermitian metric tensor (also called quantum geometric tensor) defined in the tangent subspace of a projective manifold of cell-periodic Bloch states introduced by Ahn et al.[7,23], and $\mathcal{C}_{bac}^{nm} = A_b^{nm} \mathfrak{D}_a^{nm} A_c^{mn}$ is the Hermitian connection (also known as quantum geometric connection). The prime on the summation notation means in the summation all the band indices cannot be equal. The SBCP can be cast into

$$T_{abc}^n = \mathfrak{R} \sum_m{}' \left[ \frac{i}{\omega_{nm}^2} (\mathcal{C}_{bac}^{nm} + \mathcal{C}_{abc}^{nm} - \mathcal{C}_{cba}^{mn}) - \frac{i}{\omega_{nm}^3} (\Delta_a^{nm} Q_{bc}^{nm} + \Delta_b^{nm} Q_{ac}^{nm} - \Delta_b^{mn} Q_{ca}^{mn}) \right] + TBC$$

$$= \sum_m{}' \left[ \frac{1}{\omega_{nm}^2} (\Gamma_{bac}^{nm} + \Gamma_{abc}^{nm} + \Gamma_{cba}^{nm}) - \frac{1}{2\omega_{nm}^3} \Delta_a^{nm} \Omega_{bc}^{nm} \right] + TBC,$$

where the (minus) imaginary part of quantum geometric connection $\Gamma_{abc}^{nm} = -\mathfrak{I} \mathcal{C}_{abc}^{nm}$ is known as the symplectic connection, and $\Omega_{bc}^{nm} = -2\mathfrak{I} Q_{bc}^{nm}$ is the symplectic form (Berry curvature). We have used the antisymmetric property of symplectic connection: $\Gamma_{cba}^{mn} = -\Gamma_{cba}^{nm}$. Because only the SBCP with the latter two subscripts symmetrized contributed to the physical positional shift $\delta r_a = T_{abc} E_b E_c$, here it is sufficient to retain such terms in the above equation. Then we obtain

$$T_{abc}^n = \sum_m{}' \frac{1}{\omega_{nm}^2} (\Gamma_{bac}^{nm} + \Gamma_{acb}^{nm} + \Gamma_{cba}^{nm}) + \mathfrak{R} \sum_{m,\ell}{}' \frac{A_b^{nm} A_a^{m\ell} A_c^{\ell n} + A_a^{nm} A_c^{m\ell} A_b^{\ell n} + A_c^{nm} A_b^{m\ell} A_a^{\ell n}}{\omega_{nm} \omega_{nl}}.$$

In the main text we employ this expression for the symplectic connection term and the three-band contribution.

**First-principles calculations**

Our first-principles calculations were performed using density functional theory as implemented in the Vienna Ab-initio Simulation Package[40,41] with projector augmented wave method. The exchange-correlation functional was treated using the generalized gradient approximation with the Perdew-Burke-Ernzerhof parametrization[42]. The cutoff energy was set to be 400 eV. The 3D Brillouin zone of bulk Fe$_3$GaTe$_2$ was sampled with a Γ-centered k-point grid of 15 × 15 × 3. Van der Waals interactions between adjacent layers were considered via the Grimme DFT-D3 method[43]. To calculate the nonlinear transport of bulk Fe$_3$GaTe$_2$, a Wannier tight-binding Hamiltonian was constructed using the Wannier90 package[44]. The *d*-orbitals of Fe atoms and the *p*-orbitals of Ga and Te atoms were selected as the initial projection basis. The intrinsic third-order anomalous Hall conductivity and Berry curvature quadrupole were calculated using the Wannier Hamiltonian on a dense k-mesh of 801 × 801 × 161.

**Centrosymmetric magnetic point groups compatible with intrinsic THE**

By symmetry analysis, we found that 19 centrosymmetric magnetic point groups permit intrinsic THE, including $\bar{1}$, $2/m$, $2'/m'$, $mmm$, $m'm'm$, $4/m$, $4'/m$, $4/mmm$, $4'/mm'm$, $4/mm'm'$, $\bar{3}$, $\bar{3}m$, $\bar{3}m'$, $6/m$, $6'/m'$, $6/mmm$, $6'/m'mm'$, $6/mm'm'$ and $m\bar{3}m'$. Among them, 10 are compatible with ferromagnetism, i.e., $\bar{1}$, $2/m$, $2'/m'$, $m'm'm$, $4/m$, $4/mm'm'$, $\bar{3}$, $\bar{3}m'$, $6/m$ and $6/mm'm'$. We remind that these 10 are also all the magnetic point groups that are compatible with ferromagnetism. This means that all centrosymmetric ferromagnets support the intrinsic THE.

**Excluding the Berry curvature quadrupole mechanism**

The third-order Hall conductivity contributed by Berry curvature quadrupole (BCQ) $\chi_{yxxx}^{BCQ}$ is given by $\chi_{yxxx}^{BCQ} = \frac{e^4\tau^2}{2\hbar^3} Q_{xxz}$, where $\tau$ is relaxation time, $Q_{xxz}$ is in-plane Berry curvature quadrupole $Q_{xxz} = \int [d\boldsymbol{k}] \partial_x \partial_x \Omega_z f_0$, with $\Omega_z$ the Berry curvature for a single state. To assess the relevance of the BCQ mechanism, we calculated the third-order Hall conductivity due to BCQ, and compared it with the experimental value. At T = 150 K, the calculated value is $1.34 \times 10^{-9}$ $\Omega^{-1}\text{mV}^{-2}$, where $\tau$ was estimated using Drude model, which is four orders of magnitude smaller than the experimental value. Therefore, the Berry curvature quadrupole is not relevant here. Indeed, the scaling law analysis did not find a quadratic scaling characteristic of the Berry curvature quadrupole contribution.

**Data availability**

The data that support the findings of this study are available from the corresponding author upon reasonable request.

**Acknowledgements**

The research was supported by the Ministry of Science and Technology of China (Grant No. 2024YFA1410100), the National Natural Science Foundation of China (Grant No. 12474121, 12488101), the Science and Technology Commission of Shanghai Municipality (Grant Nos. 2019SHZDZX01), the Innovation Program for Quantum Science and Technology (Grant No. 2021ZD0302500). L.L. acknowledges the Xiaomi Young Scholar program and the Yangyang Development Fund. C.X. is sponsored by the National Natural Science Foundation of China (grant No. 12574114) and the start-up funding from Fudan University. S.A.Y. is supported by The HK PolyU Start-up Grant No. (P0057929).

**Author contributions**

L.L. and Y.C.: conceived and designed the experiments; Y.C.: performed device fabrication, transport measurements, and data analysis with the assistance of Y.G., Q.S., H.C., and W.G.; R.Z., Y.Y., X.P., and J.J.: contributed to data analysis. C.X., S.A.Y., X.F., and H.L.: performed the symmetry analysis and first-principles calculation. Q.S.: performed STEM experiments. L.L., C.X., S.A.Y., Y.C., and X.F. wrote the manuscript and all authors contributed to its final version.

**Competing financial interests' statement**

The authors declare no competing financial interests.

Correspondence and requests for materials should be addressed to L.L. (liul21@sjtu.edu.cn), C.X. (congxiao@fudan.edu.cn) and S.A.Y. (shengyuan.yang@polyu.edu.hk).